# AN IMPROVED GEF FAST ADDITION ALGORITHM

*Md.Mizanur Rahaman, Md.Shahadat Hossain, Md.Rakib Hasan and M.M.A. Hashem*
*Department of Computer Science & Engineering*
*Khulna University of Engineering & Technology (KUET)*
*Khulna 9203, Bangladesh*
*E-mail: hashem@cse.kuet.ac.bd*

## ABSTRACT

In this paper, an improved GEF fast addition algorithm is proposed. The proposed algorithm reduces time and memory space. In this algorithm, carry is calculated on the basis of arrival timing of the operand's bits without overhead of sorting. Intermediate terms are generated from the most significant bit and the carry is generated from the least significant bit using the functions of efficient operators. This algorithm shows better performance for use in the fastest computational devices of the near future.

## 1. INTRODUCTION

Addition is the most fundamental operation in computation. When the computation speed of the addition increases, the performance of the computer also increases. There are many algorithms to perform fast addition. Fast addition is achieved by using logarithmic time parallel adders[1,2], carry-lookahead (CLA)[3] and conditional sum adder (CSMA)[4,5] based algorithms. Among these wide varieties of algorithms, Carry propagation addition (CPA)[6] has been used in various applications. The CLA and CSMA consider the special case where all the inputs arrive simultaneously. Generalized Earliest-First (GEF)[7] Fast addition algorithm performs the best. It uses advance carry propagation technique, which makes it faster than the existing algorithms. GEF fast addition algorithms consider that the inputs don't arrive simultaneously. It processes two-bit block at each iteration. It generates carry from the earliest arriving signals without having to process least significant bits first. The GEF use sorting and searching to build carry. It also uses extra memory. This paper proposes an improved GEF fast addition algorithm based on the bit level operation. It reduces the processing steps and makes the addition faster in comparison with other algorithms. The improved algorithm shows good performance in terms of area and speed than GEF fast addition algorithm. The algorithm performs addition by generating carry from LSB directly and combines the other bits to form intermediate terms. The algorithm also saves memory space. The way of its calculation is parallel in nature and it is faster than the existing algorithms.

## 2. PREVIOUS WORK

Among the various addition algorithms CLA, CSMA, Dual forward prediction (DFP)[7] algorithm and GEF fast addition algorithm are the most popular algorithms. The following equations can be used to compute sum and carry.

$$s_i = p_i \oplus c_{i-1} \qquad (1)$$

$$c_i = c_{i-1}.r_i + c_{i-1}.g_i \qquad (2)$$

$$c_i = g_i + p_i.c_{i-1} \qquad (3)$$

$$c_i = g_i + r_i.c_{i-1} \qquad (4)$$

Where the $p_i$, $g_i$ and $r_i$ are defined as $p_i = a_i \oplus b_i$, $g_i = a_i \cdot b_i$ and $r_i = a_i + b_i$ respectively. Note that the initial carry $c_0 = g_0$ and the subscript 'i' denotes the bit position starting from LSB. Eq (1) is for sum and Eq (2) is for carry of CSMA. Eq (3) and (4) both used to generate carry for CLA. We use Eq (3) for "CLA_I" and Eq (4) for "CLA_II". The DFP algorithm improves the adder performance using Delay Profile (DP), but it does not guarantee the optimality of the performance in any way and it only considers the next two bits at each iteration. DP is an array consisting of N elements of maximum arrival timing of corresponding augend and addend bits.

## 3. GEF FAST ADDITION ALGORITHM

The GEF algorithm, which considers the order of computation at all possible bit positions in each iteration. The basic idea of the GEF addition algorithm is based on two facts. *Firstly,* when some bits arrive earlier than others, it combines them first to generate intermediate term ($g_i$, $r_i$) to improve the performance. The generated intermediate terms, i.e. ($g_i$, $r_i$) can be treated as primitive terms. *Secondly,* the combination does not have to start from LSB it does not violate associative and non-commutative properties. The bits are adjacent when there are no other bits between them. The GEF algorithm processes two elements at each iteration. It sorts the arrival time in ascending order and transfers to another array. The new array is sorted with respect to bit position. Finally it combines the adjacent bits and generates carry if previous carry is available according to Fig: 1.

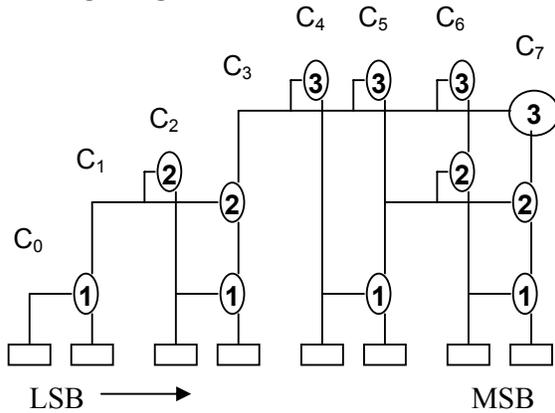

**Fig. 1** The generation of carry signals for the GEF based adder. Each circle represents a ternary operator and the number inside the circle shows the timing normalized with respect to the operator.

The GEF algorithm has five steps for N bit addition:
**Step-1: Setup:** Initialize an array P_list with N elements,
For $0 \leq i \leq N\text{-}1$, let P_list[i] = DP[i]. Each entry in P_list consists of the arrival timing and the bit position. Initialize T_list with zero elements and the structure of T_list is same as P_list.
**Step-2: Generate new T_list:** Sort P_list in ascending order according to the arrival timing. Move the elements that are equivalent to P_list [0] from P_list to T_list. Sort T_list in ascending order according to the bit position. Retain the bit position information.
**Step-3: Combine adjacent bits:** In T_list, if the bits are adjacent to each other, combines them with ternary operator. Insert the generated new terms with bit positions and timings back into P_ list.
**Step-4: Repeat.** Repeat steps 2 & 3 until only one element left in P list.
**Step-5: Generate** sum signals and finish.

## 4. PROPOSED GEF FAST ADDITION ALGORITHM

The basic idea of the proposed GEF fast addition algorithm is based on three facts. *Firstly*, the P_time [i] is an array of arrival time of the operands. It does not need to be sorted. There is no bit position array so memory is saved and it speedup the task. *Secondly,* when some bits arrive earlier than the others, it calculates carry signal from LSB immediately, otherwise it calculates intermediate term ($g_i$, $r_i$). *Thirdly,* it combines two or more terms at a time if they are adjacent. The computation will be performed randomly prior to the arrival time. It combines two adjacent terms using any one of the ternary operators. Three ternary operators "$\otimes$", "$\nabla$", "$\Delta$" are defined as

$a.b + a.c = \overline{a} \otimes (b, c)$         (5)
$b + (c.a) = a \nabla (b,c) = c \nabla (b,a)$     (6)
$b \oplus (c.a) = a \Delta (b,c) = c \Delta (b,a)$    (7)

If the two ($g_i$, $r_i$) are adjacent then it calculates the new ($g_i$, $r_i$) according to the Eq (8), (9) & (10).

$$(a,b) = (\alpha,\beta) \otimes (\gamma,\delta) \Rightarrow \begin{cases} a = \alpha \otimes (\gamma,\delta) \\ b = \beta \otimes (\gamma,\delta) \end{cases} \dots(8)$$

$$(a,b) = (\alpha,\beta) \nabla (\gamma,\delta) \Rightarrow \begin{cases} a = \alpha \nabla (\gamma,\delta) \\ b = \beta.\delta \end{cases} \dots(9)$$

$$(a,b) = (\alpha,\beta) \Delta (\gamma,\delta) \Rightarrow \begin{cases} a = \alpha \Delta (\gamma,\delta) \\ b = \beta.\delta \end{cases} \dots(10)$$

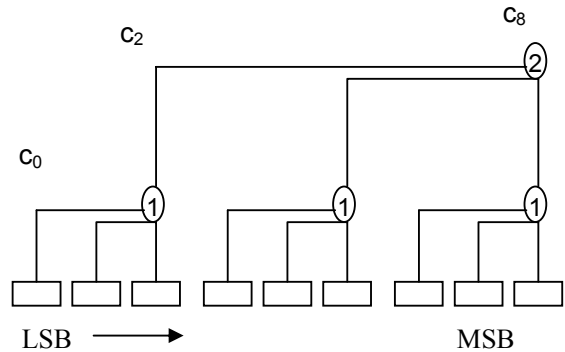

**Fig. 2(a)** The basic figure of proposed IGEF algorithm, which contains carries only $3^i$.

### 4.1 Steps of proposed GEF addition algorithm:

**Step-1:** P_time[i] is a single dimensional array. Initialize the array with maximum arrival timing of the corresponding bit of the two operands for $0 \leq i \leq N-1$. T_time[i] is an array initialized with Zero elements.

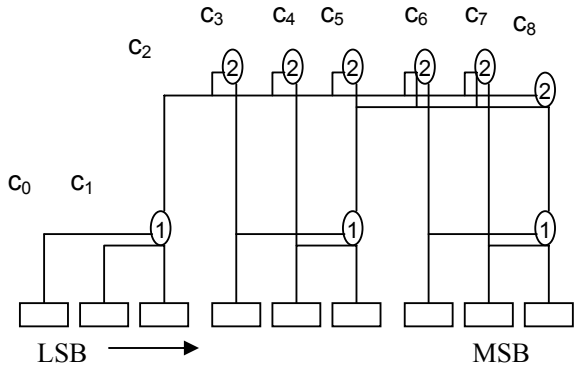

**Fig. 2(b)** The generation of carry signals for the proposed IGEF based adder. Each circle represents a ternary operator and the number inside the circle shows the timing normalized with respect to the operator. A detail of the basic figure, which contains all, carries Like $C_0 \ldots C_{26}$.

**Step-2:** Generate T_time[i]: Search the minimum value from P_time[i] in each iteration. Transfers all the value which is equivalent to minimum value of P_time[i] to T_time[i] in corresponding bit position and NULL the corresponding position of P_time[i].

**Step-3:** Find adjacent value in T_time[i]. If found, generate carry signal from LSB otherwise combine them to generate intermediate term ($g_i$, $r_i$). The new generated term not is transfer to P_time.

**Step-4:** Repeat steps 2 & 3 until rest of one element in P_time[i].

**Step-5:** Finally generate sum signal.

The algorithm guarantees the optimality in each step and hence, the generated adder structure is optimized in terms of speed and memory. Note that two terms are considered as adjacent to each other when there is no other term between them. They can have inconsecutive bit positions. If there are three adjacent bits in the T-time[i], it cascades two ternary operators to compute the 3-bit block. The arrival time will be transferred from DP[i] to P_time[i] to initiate the array. The T-time[i] is initiated by NULL value. There need not sort the P_time[i]. In first step, it just searches the P_time[i] for minimum value. In Second step, it generates a new array T_time and then transfers the value equivalent to minimum value into T_time from P_time[i]. GEF algorithm needs to sort the N value with respect to arrival time and transferred the value equivalent to P_time[0] from P_time[i] to T_time[i].

## 5. TIMING ANALYSIS

Since the equations for sum path and carry path are independent, it can be analyzed them separately. The carry-generation equations are repeated below for convenience.

For CSMA and CCA
$$c_i = c_0 \otimes (r_1, g_1) \otimes \ldots \otimes (r_{i-1}, g_{i-1}) \otimes (r_i, g_i) \quad (11)$$

For ELMA and CLA_I
$$c_i = c_0 \nabla (g_1, p_1) \nabla \ldots \nabla (g_{i-1}, p_{i-1}) \nabla (g_i, p_i) \quad (12)$$

For CLA_II
$$c_i = c_0 \nabla (g_1, r_1) \nabla \ldots \nabla (g_{i-1}, r_{i-1}) \nabla (g_i, r_i) \quad (13)$$

Assume that all the inputs arrive simultaneously and two bits are combined at each step using a ternary operator, i.e., r =2. According to the associative property, it combines any two terms in the equations in any order. Hence, the performance will be determined by the latency of each operator and the time to setup $g_i$, $r_i$ and $p_i$ terms. Since the latency of the generation of $p_i$ is longer than those of $r_i$ or $g_i$. By considering the setup time and the latency of the operators, it concludes that the fastest way to generate carry is via Eq (13). Let's now consider the generation of sum signal. The equations for sum generation are given below.

For CCA
$$s_i = p_i \oplus [c_0 \otimes (r_1, g_1) \otimes \ldots \otimes (r_{i-1}, g_{i-1}) \otimes (r_i, g_i)] \quad (14)$$

For CSMA
$$s_i = c_0 \otimes (r_1, g_1) \otimes \ldots \otimes (r_{i-1}, g_{i-1}) \otimes (r_{i-2}, g_{i-2})$$
$$\otimes (p_i \oplus r_{i-1}, p_i \oplus g_{i-1}) \quad (15)$$

For CLA_I
$$s_i = p_i \oplus [c_0 \nabla (g_1, p_1) \nabla \ldots \nabla (g_{i-1}, p_{i-1}) \nabla (r_i, g_i)] \quad (16)$$

For CLA_II
$$s_i = p_i \oplus [c_0 \nabla (g_1, r_1) \nabla \ldots \nabla (g_{i-1}, r_{i-1})] \quad (17)$$

According to these equations, it depicts the generation of the last sum bit for four of the adder types for r=2. For CLA_II and CCA adder types, the $P_i$ is always merged at the last step. When the word length N, is even, the ($g_{L-2}, r_{L-2}$) term can only be combined with the other terms at the second level, which causes inefficiency. In contrast, for CSMA and the ELMA, the ($g_{L-2}, r_{L-2}$) term can always be combined with $p_i$. Hence, unless N happens to be (2N+1), CSMA will always require one less operator on the critical path than either CLA_II. The blocking factor, which means the number of bits can be processed in one logic gate, is denoted by r. When N is greater than eight, the number of ternary operator will be greater than three on the critical path. The CLA II will be faster than CSMA, CCA and CLA_I. Among these algorithms the GEF algorithm is faster. The proposed IGEF algorithm is faster and reduces the memory space than GEF algorithm. Table 1 & 2 represent the timing performance of GEF & IGEF algorithms.

**Table 1:** Example for GEF algorithm

| It | Li | Bit Position | | | | | | | | | | | |
|---|---|---|---|---|---|---|---|---|---|---|---|---|---|
|   |   | 0 | 1 | 2 | 3 | 4 | 5 | 6 | 7 | 8 | 9 | 10 | 11 |
| 0 | P |   |   | 2 | 2 | 3 | 3 | 4 | 5 | 4 | 3 | 2 |   |
|   | T | 0 | 1 |   |   |   |   |   |   |   |   |   | 1 |
| 1 | P |   |   |   |   | 3 | 3 | 4 | 5 | 4 | 3 |   |   |
|   | T |   | 2 | 2 | 2 |   |   |   |   |   |   | 2 | 1 |
| 2 | P |   |   |   |   |   |   | 4 | 5 | 4 |   |   |   |
|   | T |   | 2 |   | 3 | 3 | 3 |   |   |   | 3 |   | 3 |
| 3 | P |   |   |   |   |   |   |   | 5 |   |   |   |   |
|   | T |   |   |   | 4 |   | 4 | 4 |   | 4 |   |   | 4 |
| 4 | P |   |   |   |   |   |   |   |   |   |   |   |   |
|   | T |   |   |   | 4 |   |   | 5 | 5 |   |   |   | 5 |
| 5 | P |   |   |   |   |   |   |   |   |   |   |   |   |
|   | T |   |   |   |   |   |   | 6 |   |   |   |   | 6 |
| 6 | P |   |   |   |   |   |   |   |   |   |   |   |   |
|   | T |   |   |   |   |   |   |   |   |   |   |   | 7 |

**Table 2:** Example of proposed IGEF algorithm

| It | Li | Bit Position | | | | | | | | | | | |
|---|---|---|---|---|---|---|---|---|---|---|---|---|---|
|   |   | 0 | 1 | 2 | 3 | 4 | 5 | 6 | 7 | 8 | 9 | 10 | 11 |
| 0 | P |   | 1 | 2 | 2 | 3 | 3 | 4 | 5 | 4 | 3 | 2 | 1 |
|   | T | 0 |   |   |   |   |   |   |   |   |   |   |   |
| 1 | P |   |   | 2 | 2 | 3 | 3 | 4 | 5 | 4 | 3 | 2 |   |
|   | T | 0 | 1 |   |   |   |   |   |   |   |   |   | 1 |
| 2 | P |   |   |   |   | 3 | 3 | 4 | 5 | 4 | 3 |   |   |
|   | T |   | 2 | 2 | 2 |   |   |   |   |   |   | 2 | 1 |
| 3 | P |   |   |   |   |   |   | 4 | 5 | 4 |   |   |   |
|   | T |   |   |   | 3 | 3 | 3 |   |   |   | 3 |   | 3 |
| 4 | P |   |   |   |   |   |   |   | 5 |   |   |   |   |
|   | T |   |   |   |   | 4 | 4 |   | 4 |   |   |   | 4 |
| 5 | P |   |   |   |   |   |   |   |   |   |   |   |   |
|   | T |   |   |   |   |   |   | 5 | 5 |   |   |   | 5 |
| 6 | P |   |   |   |   |   |   |   |   |   |   |   |   |
|   | T |   |   |   |   |   |   |   |   |   |   |   | 6 |

## 6. CONCLUSION

In this paper, we have proposed some logic for removing limitations of the GEF fast addition algorithm. The proposed algorithm is based on parallel processing on hardware level. The algorithm is developed to solve the problems of existing addition algorithms by considering global timing properties and multiple bits in each iteration. The GEF fast algorithm has sorting complexity. So timing performance is not good enough and more memory is required for maintaining bit position and arrival time. In the proposed IGEF fast algorithm, the sorting complexity is removed by finding minimum arrival timing in each iteration. This technique reduces the time and use of memory. The algorithm uses the appropriate operator, which also reduces the operational time. Hence, the proposed algorithm is optimal with respect to time than the GEF and other addition algorithms.